\documentclass{article}
\usepackage{url}
\usepackage{times}
\usepackage{gensymb}
\usepackage{graphicx} 
\usepackage{subfigure} 

\usepackage{natbib}

\usepackage{algorithm}
\usepackage{algorithmic}

\usepackage[accepted]{icml2017}

\icmltitlerunning{Ephemeral Recommendations}

\begin{document} 

\twocolumn[
\icmltitle{Ephemeral Context to Support Robust and Diverse Music Recommendations}






\begin{icmlauthorlist}
\icmlauthor{Pavel Kucherbaev}{df}
\icmlauthor{Nava Tintarev}{df}
\icmlauthor{Carlos Rodriguez}{sw}
\end{icmlauthorlist}

\icmlaffiliation{df}{Delft University of Technology, Delft, The Netherlands}
\icmlaffiliation{sw}{University of New South Wales, Sydney, Australia}

\icmlcorrespondingauthor{Pavel Kucherbaev}{p.kucherbaev@tudelft.nl}

\icmlkeywords{music discovery, life-logging, context-awareness}

\vskip 0.3in
]



\printAffiliationsAndNotice{\icmlEqualContribution} 

\begin{abstract} 
While prior work on context-based music recommendation focused on fixed set of contexts (e.g. \emph{walking}, \emph{driving}, \emph{jogging}), we propose to use multiple sensors and external data sources to describe momentary (ephemeral) context in a rich way with a very large number of possible states (e.g. \emph{jogging fast along in downtown of Sydney under a heavy rain at night being tired and angry}). With our approach, we address the problems which current approaches face: 1) a limited ability to infer context from missing or faulty sensor data; 2) an inability to use contextual information to support novel content discovery. 


\end{abstract} 

\section{Introduction}


Imagine the following persona: Anna is a university student in graphic design. She is active, easy going, and organized; keeping a structured well-planned calendar. She is 21 and high on the trait of Openness to Experience \cite{tkalcic2015personality}, and enjoys traveling to new places and meeting new people, as well as discovering and listening to new music using online radio services.\\
To improve Anna's listening experience, we consider her context. To decide which aspects of context to look at, we ran an exploratory crowdsourced survey. Out of 103 respondents 97 listen to different music based on their mood (e.g., happy, calm, sad), 92 based on their activity (e.g., commuting, jogging); 38 based on the ambience (e.g. sunny, rainy, loud, quiet), and 32 based on the location (e.g., a city park, a beach). From free-text answers we know that preferences of respondents also change according to the weather, time of day, people around, headphones or speakers used, upcoming concerts, the difficulty of work they do, languages they learn, reminiscence, and music that they just heard somewhere. Several respondents suggested combined causes, such as friends that are around and the activity they performed together.

We propose that Anna's online radio would suggest unexpected pleasant surprises based on her momentary unique (and therefore ephemeral) context. The contribution of this paper is using multiple sensors and external data sources to make the inference of life-logging events richer (through combinations of inputs), and more reliable (using multiple sensors to improve fault tolerance). It describes how rich context information can be combined in a large number of ways to improve the diversity of recommendations, which will lead to more opportunities for music discovery. 



\section{Related work}


The type of contextual recommendations that can be made is shaped by sensors and signal processing used. Nowadays it is possible to accurately detect activities such as biking, driving, running, or walking based on smartphone sensors \cite{Liao2015}, or based on environmental sound cues \cite{Shaikh2008}. It is also possible to detect personality traits based on phone call patterns and social network data of the user \cite{deOliveira2011}. Similarly, interest in an object can be inferred based on ambient noise levels, and positions of people and objects in relation to each other \cite{dim2015automatic}. In the SenSay system, phone settings and preferences are set based on detected environmental and physiological states \cite{Siewiorek2003}.

With improvements in smartphone technology, there is a lot of potential for using rich contextual information to improve recommendations, in particular considering that people prefer to listen to different music in different contexts \cite{Cunningham08, Schedl2014, Su2010}. Among the first to propose a context-aware music recommendation system are Park et al. \cite{Park2006}. They used weather data (from sensors and external data sources), and user information, to predict the appropriate music genre, tempo, and mood. Music can also be recommended based on user's heart beat to bring its rate to a normal level \cite{Liu2009}; activities detected automatically (e.g. running, walking, sleeping, working, studying, and shopping) \cite{Wang2012}; driving style, road type, landscape, sleepiness, traffic conditions, mood, weather, natural phenomena \cite{Baltrunas2011}; and emotional state to help to transition to a desired state \cite{Han2010}. Soundtracks have also been recommended for smartphone videos based on location (using GPS and compass data for orientation), and extra information from 3rd party services such as Foursquare \cite{Yu2012}.

These examples use sensors and external data sources for music recommendation. Some of these context-aware music discovery systems recommend not just relevant, but new music to users \cite{Wang2013}. Our contribution is to combine rich context in a way that is a) fault tolerant, and b) aims to facilitate music discovery, by constructing a momentary ephemeral context. 


\section{Approach}
\begin{figure}[t]
\centering
\includegraphics[width=\columnwidth]{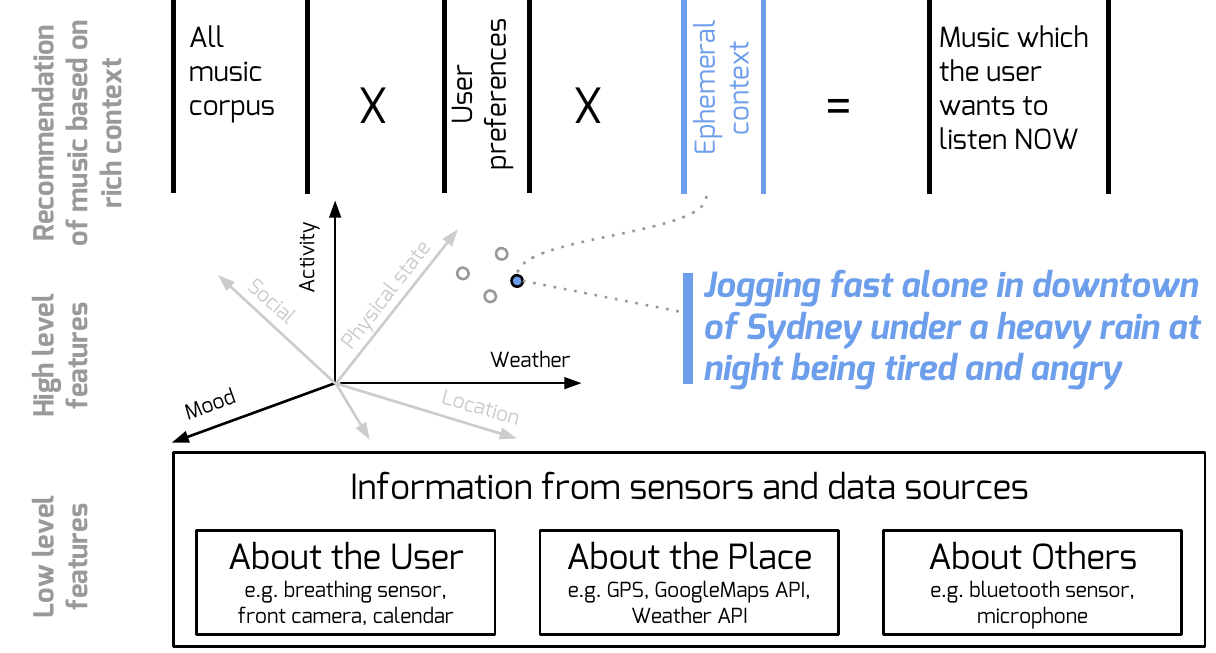}
\caption{An example of music recommendation using as input ephemeral context constructed from high-level features inferred from smartphone sensors, wearables, and external data sources.}
\label{framework}
\end{figure}




Here we give an example of how our approach could work to recommend music to Anna by detecting her ephemeral context based on high-level features (e.g. activity, mood, or the weather), which are inferred from low-level sensor data (see Figure \ref{framework}) and discuss the benefits of such approach.


\textbf{From low-level to high-level features.} We infer that Anna's \emph{activity} is ``jogging'' from the pattern of her smartphone's accelerometer and GPS, and because this activity was also planned in her calendar. Her \emph{speed} is classified as ``fast'' for jogging, because her speed is 15km/h, while usually she runs 13km/h. For the \emph{social} component she is classified as ``alone'', since her Bluetooth sensor does not see Bluetooth sensors of her friends' smartphones and the microphone does not recognise voices around. The \emph{location} is ``downtown of Sydney'' based on the coordinates given by her smartphone's GPS, and the point of interest identified by GoogleMaps API, as well as recent reviews about it from FourSquare. The \emph{weather} is ``heavy rain'' according to the moisture sensor of her phone and the weather forecast for the location from Weather API. The \emph{time of day} is ``night'', because her smartphone time is 23:56. Her \emph{physical state} is ``tired'' based on the high heart rate measured by her smart bracelet, and the respiratory pattern coming from her breathing sensor. Her \emph{mood} is detected as ``angry'' by her smartphone front camera \cite{Busso2004} and based on her public interactions on social media (e.g., angry emoticon). We combine these high-level features to construct a momentary ephemeral context, which becomes: ``\emph{Jogging fast alone in downtown of Sydney under a heavy rain at night being tired and angry}".

\textbf{From individual recommenders to a hybrid one.} We propose to use several individual recommenders focused on different sets of high-level features (e.g. a recommender looking only at location, weather, and time). A hybrid recommender later weights recommendendations of each individual one, based on explicit preferences of Anna, and the reliability of underlying high-level features, if detected at all. Anna can change weights to make an emphasis on a certain aspect, such as location or activity, depending on the way she wants to explore music. We provide an interactive web-based demonstration\footnote{The demo, the code, and the results of the exploratory survey are available at \url{https://github.com/pavelk2/ICML2017Demo}.} of how such a hybrid recommender might work based on ephemeral context.



\textbf{Benefits.}
Our approach allows us to effectively address fault tolerance and leverage music discovery:

\textbf{Fault tolerance.} Different factors are used as a measure of fault tolerance. For example, if GPS and calendar locations are different, the system will omit location-based recommendations from the hybrid recommender.

\textbf{Music discovery.} Since ephemeral context frequently changes, the recommendations supplied will vary from moment to moment, leading to more opportunities for music discovery (e.g. 8 high-level features taking 8 values each, give more than 16mln combinations). 

\section{Outlook}

Our next steps will be dedicated to \emph{identification of combinations of high-level features} influencing music preferences, possibly via a user study; to \emph{evaluation of recommendations' relevance} with cultural preferences in mind, potentially using crowdsourcing \cite{Ahn2008}, and to study how to make such rich user profiling compliant with \emph{privacy} concerns \cite{coles2011looking}. We also plan to study how to improve the transparency of ephemermal recommendations using textual and visual explanations. As such, we aim to deliver a streaming music experience, driven by context, while giving the user a sense of transparency and control.
We are confident that music discovery through rich context is a very promising research topic, allowing streaming services to provide better personalized experiences to their listeners.


 



\clearpage
\bibliography{references}
\bibliographystyle{icml2017}

\end{document}